\documentclass[prb,twocolumn,showpacs,preprintnumbers,amsmath,amssymb]{revtex4-1}
\usepackage{graphicx}
\usepackage{epstopdf}

\begin{document} 
\title{Metal-superconductor transition in two-dimensional electron systems with
fractal-like mesoscopic disorder} 
\author{D. Bucheli$^1$, S. Caprara$^{1,2}$, C. Castellani$^{1,2}$, and 
M. Grilli$^{1,2}$} 
\affiliation{$^1$CNISM and Dipartimento di Fisica Universit\`a di Roma ``La Sapienza'', 
piazzale Aldo Moro 5, I-00185 Roma, Italy\\
$^2$Consiglio Nazionale delle Ricerche, Istituto dei Sistemi Complessi, 
via dei Taurini, I-00185 Roma, Italy}   
\begin{abstract} 
Motivated by recent experimental data on thin film superconductors and oxide interfaces we propose 
a random-resistor network apt to describe the occurrence of a metal-superconductor transition 
in a two-dimensional electron system with disorder on the mesoscopic scale.
We explore the interplay between the 
statistical distribution of local critical temperatures and the occurrence of a lower-dimensional 
(e.g., fractal-like) structure of a superconducting cluster embedded in the two-dimensional
network. The thermal evolution of the resistivity is determined by an exact calculation and, for comparison, 
a mean-field approach called effective medium theory (EMT).
Our calculations reveal the relevance of the distribution of critical temperatures for clusters with low
connectivity. In addition, we show that the presence of spatial correlations requires a modification
of standard EMT to give qualitative agreement with the exact results.



\end{abstract}   
\date{\today} 
\maketitle 
   
\section{Introduction}\label{intro}
Although two-dimensional (2D) electron systems have always been a topical subject in condensed 
matter physics, further renewed interest in this field has been triggered by the recent discovery 
of superconductivity at the metallic interface between two insulating oxide 
layers.\cite{reyren,triscone,espci} In addition, experiments on certain thin conventional 
superconducting films\cite{sacepe1,sacepe2,sacepe3,mondal} reveal pseudogap effects, a phenomenon 
which could unveil new features of the superconducting transition in low-dimensional disordered 
systems.\cite{review_feigelman} In the particularly interesting cases of homogeneously disordered 
superconducting titanium nitride or indium oxyde thin films, scanning tunneling spectroscopy data 
display inhomogeneities in the local density of states on mesoscopic spatial scales,
clearly indicating 
the existence of an inhomogeneous superconducting state. Since tunneling experiments cannot 
directly access the 2D metallic layers at the oxide interfaces (although some attempts made on samples 
with very few top layers have detected interesting inhomogeneous textures\cite{salluzzo}) 
magnetization and transport measurements are the only way to investigate these systems.
Also in this case, evidences for an electronic phase separation have been found in 
LaAlO$_3$/SrTiO$_3$ layers.\cite{ariando} Moreover, the sheet resistance curves $R_\square (T)$ near 
$T_c$, both in LaAlO$_3$/SrTiO$_3$ [\onlinecite{triscone}] and LaTiO$_3$/SrTiO$_3$ 
[\onlinecite{espci}] systems, show a peculiar feature of a marked tail on the low-temperature side. 

Despite their differences, oxide interfaces and thin films share the common feature of a broad 
superconducting transition. In a previous work,\cite{CGBC} we investigated the possible origin of 
this pronounced width and found that superconducting fluctuations (alone) cannot account for its
occurrence, while a model of quenched mesoscopic disorder can well produce the observed broad 
transitions. In this previous work we also found that ``tailish'' features can occur in the 
resistivity curves when disorder has spatial correlations. Specifically, while uncorrelated disorder 
can generically account for the pronounced width of the metal-superconductor transition, tails only 
occur when disorder displays a correlated character or when it occurs on nearly one-dimensional subsets 
of the system. It is therefore important to understand how (correlated) disorder can act as to give 
rise to the tailish resistivity 
because this feature can be a signature of 
specific physical mechanisms at work in these systems. One natural possibility is that disorder 
correlations and/or nearly one-dimensional (or fractal) structures arise in the system from 
nanofractures, topological defects like crystal dislocations, steps, twinning domains, and so on. 
Another possibility is that low-dimensional superconducting subsets can spontaneously arise in the
electronic gas even in the presence of a homogeneously distributed disorder. In particular, one 
should remember that, besides fluctuations of the energy gap $\Delta$, disorder can cause the 
localization of electrons, transforming an otherwise metallic system into an insulator. If the metal 
is also a superconductor then, at low temperatures, disorder can induce a superconductor-insulator 
transition.\cite{leema,feigelman} In this case, a theoretical proposal has been put forward, which 
predicts that superconductivity at high disorder is maintained by a fragile coherence within a small 
set of preformed Cooper pairs that are characterized by an anomalously large binding energy. 
Consequently, in the vicinity of the superconductor-insulator transition, both the insulator and 
the superconductor contain these preformed Cooper pairs that either localize, leading to an insulating 
state, or condense into a coherent state with a glassy behavior around the 
transition.\cite{leema,feigelman,ioffe,feigelman10} The superconducting state then occurs on a 
small (possibly fractal) subset of the system, which forms in the electron gas around $T_c$. 

To shed light on the physics of disordered low-dimensional superconductors, we carry out a systematic 
investigation of a 2D metallic system with quenched disorder represented by a random-resistor network 
(RRN). As in our previous work of Ref. [\onlinecite{CGBC}], the resistors represent mesoscopic 
metallic islands where superconductivity occurs at a local critical temperature $T_c$. The 
superconducting regions are assumed to be large enough to have a fully established local coherence 
and to make the charging effects negligible. As a consequence, a mutual phase coherence 
immediately establishes between two neighboring superconducting islands, as soon as both 
have become superconducting. This clearly distinguishes our framework from the case of granular 
superconductors, where the grains have usually nanoscopic sizes of the order of the coherence length 
of the pure system. In this way we disregard phase-fluctuation effects in order to unambiguously 
identify the effects of disorder (and of its specific character). In particular, to understand the 
effects of spatial correlations, we investigate here the effects of two fractal-like distributions 
of the disordered resistors. In this way, while completely disregarding the underlying mechanisms 
leading to such correlated disordered structures, we focus on their phenomenology.

The paper is structured as follows. In Section II we introduce the generic aspects of our model, 
while in Section III we report the results of our numerical calculations for various kinds of 
fractal structures and of disorder distributions. Section IV contains a discussion of the results and 
our concluding remarks.

\section{The Random-Resistor Model and Effective Medium Theory}\label{RRN}

In a previous work\cite{CGBC} we considered a model of a 2D electron gas with mesoscopic defects 
as a square lattice whose bonds are assigned a random resistivity $\rho_i$. More precisely, 
each bond is assigned a local superconducting transition temperature 
$T_c^{(i)}$, extracted from a given distribution $\mathcal W(T_c)$, and the resistivity of the 
bond is written as $\rho_i=\rho_0\theta(T-T_c^{(i)})$, with the same high-temperature value $\rho_0$ 
on all bonds, $\theta(x)$ being the Heaviside step function. By decreasing the temperature $T$, 
more and more bonds become superconducting, and global superconductivity establishes as soon as a 
percolating superconducting path is formed in the system.

The simplest description of a RRN is provided by the effective medium 
theory\cite{effectivemedium,kirkpatrick} (EMT), which is a mean-field treatment replacing the random resistors $\rho_i$ with an effective medium resistivity $\rho_{em}$ such that
\begin{equation}
\sum_i \frac{\rho_i-\rho_{em}}{\alpha\rho_i+\rho_{em}}=0,
\label{effmed}
\end{equation} 
where the parameter $\alpha$ is related to the connectivity of the network. For 
a cubic network in $D$ spatial dimensions, $\alpha=D-1$.

In the previous work of Ref. [\onlinecite{CGBC}] (to which we refer the reader for further detail), 
we used EMT as a benchmark and compared $\rho_{em}$ with the exact numerical determination of the 
resistance of the RRN, obtained solving the Kirchhoff equations for the network in the presence of 
a difference of potential $V$ between two opposite sides of a $N\times N$ square lattice. Once the 
current $I(T)$ flowing through the network at a temperature $T$ is obtained, the resistance is found 
as $R(T)=V/I(T)$. We showed that EMT provides a very good description of the temperature dependence 
of the resistivity, independently of the distribution of $T_c$, provided disorder is spatially 
uncorrelated.

The solution of Eq. (\ref{effmed}), specialized to our case, reads\cite{CGBC,kirkpatrick} 
\begin{equation}
\rho_{em}(T)=(1+\alpha)\rho_0\theta(T-T_\alpha)\int_{T_\alpha}^{T}
dT_c \,\mathcal W(T_c).
\label{solution}
\end{equation} 
Here, $T_\alpha$ is the critical temperature of the 
effective medium and is determined by the equation 
\begin{equation}
w_s(T_\alpha)\equiv\int_{T_\alpha}^{+\infty}dT_c \,\mathcal W(T_c)=\frac{1}{1+\alpha},
\label{w_s}
\end{equation}
where $w_s(T)\equiv\int_T^{+\infty}dT_c \,\mathcal W(T_c)$ is the statistical 
weight of the superconducting bonds at a temperature $T$, measuring the frequency 
of occurrence of bonds with $T_c >T$. From Eq. (\ref{solution}) it is readily 
seen that $\rho_{em}\to\rho_0$ for $T\to\infty$ and $\rho_{em}(T_\alpha)=0$.

In the present paper we consider a model in which the superconducting bonds only form a 
spatially correlated subset of the whole system. This subset provides a ``skeleton'' 
network of bonds which become superconducting below a random local $T_c^{(i)}$, while the 
embedding system is  a metal with temperature-independent resistive bonds. 
We shall explore the joint effects of the connectivity of the network 
and of the statistical distribution of the local critical temperature $\mathcal W(T_c)$. 
We shall mainly rely on the exact numerical solution of the Kirchhoff equations, but shall refer 
to the EMT results to provide an interpretation to the outcomes of our calculations.
When dealing with EMT, the presence of spatial correlations requires the introduction of an \emph{effective} 
connectivity $\alpha$, different from the standard connectivity.
 
\section{Effects of fractal-like disordered structures}   
\subsection{The general framework}\label{framework}
To systematically investigate the effects of correlated disorder, we consider 
the occurrence of a cluster where the zero-resistance state is carried by only 
a minor set of superconducting regions, reducing the effective dimension of the 2D 
electron gas. We disregard the physical origin (classical, as due to mechanical 
stresses, non-uniform growth and so on, or quantum-mechanical, as - for example - associated with  
the occurrence of coherence on a thin cluster of Cooper pairs near the 
superconductor-insulator transition) of this low-dimensional superconducting 
``skeleton''  and we translate it into the framework of the RRN.
Specifically, we consider a RRN on which fractal-like structures are 
superimposed, i.e., clusters exhibiting strong spatial correlations and scale 
invariance (although the scale invariance is limited, once the structure
is implemented on a finite - e.g., $100\times100$ - lattice). 
Each bond of these spatially correlated clusters is given a critical temperature 
extracted from a probability distribution $\mathcal{W}(T_c)$, while the
bonds not belonging to the ``fractal'' cluster have fixed resistances $\rho_0$ 
(throughout the paper we take $\rho_0=1$). 
By varying the geometry of the 
clusters, as well as the probability distribution, the influence of these two 
aspects on the temperature dependence of the resistivity is studied. Note that in 
this approach the spatial distribution of superconducting bonds is considered to 
be independent from the probability distribution of the critical temperatures. This 
is a simplifying assumption allowing for a more general and systematic analysis of 
the separate effects of geometry and of statistical distribution.  
This assumption should of course be dropped whenever the physical mechanisms 
selecting the ``fractal'' superconducting cluster also determine its $T_c$ 
distribution.

It is worth pointing out that in the previously considered models of uniform uncorrelated
disorder\cite{CGBC}, global superconductivity occurred as a true percolative phenomenon:
Upon lowering the temperature, more and more bonds became superconducting until a
percolative cluster of superconducting bonds was reached. The geometrical support
 and the general features of this transition were typical of the percolation 
transition. On the other hand, in the present model the geometric support of bonds which 
can become superconducting is provided by a predetermined fractal-like subset of the network.
When $T$ is lowered, more and more bonds on this fractal support become superconducting
and global superconductivity only occurs when a percolating path through the whole system 
is formed inside the subset. In this case, for instance, the fractal dimension of the
superconducting support is not due to pure percolation, but it clearly depends on the dimensionality
of the supporting fractal-like subset. This ``percolation inside a fractal" is what gives rise to global
superconductivity in the present model.

\begin{figure}[h]
\begin{center}
\includegraphics[scale=0.21]{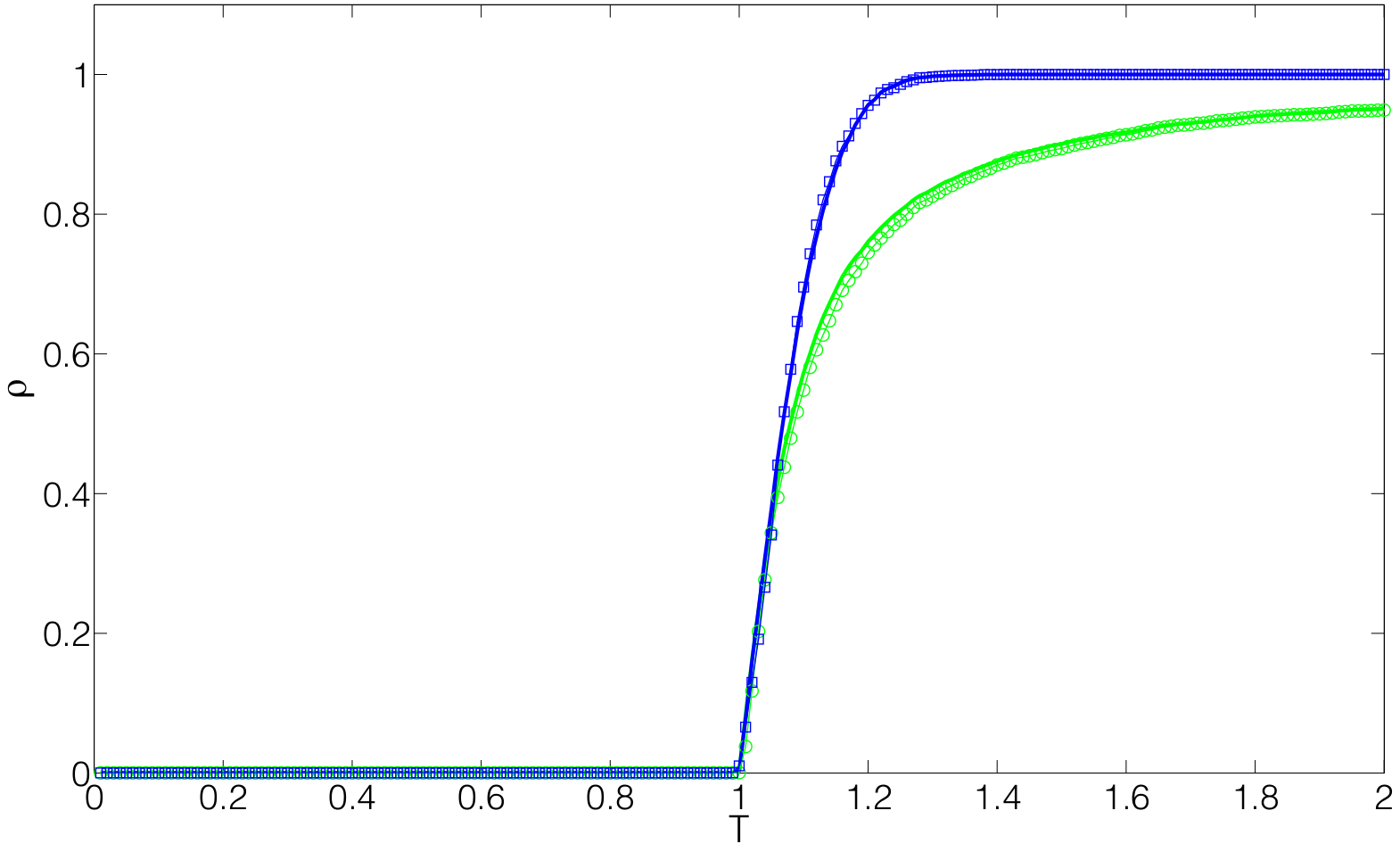}
\caption{(Color online)  Resistance curves of the uniform cluster. 
The blue curves correspond to a Gaussian distribution with $\mu=1,\sigma=0.1$
and the green to a Cauchy distribution with $\mu=1,\gamma=0.08$. 
The markers report the thermal evolution of the resistance obtained 
through the exact numerical calculations and the full lines the EMT solution.}
\label{UniformComparison}
\end{center}
\end{figure}

In order to cover a wide range of possible $T_c$ distributions we choose the two 
extreme cases of Gaussian and Cauchy statistics. The choice is motivated by their 
different asymptotic behavior: While the first has small tails and well defined 
momenta at all orders, the second is characterized by strong tails causing all 
momenta except the mean to diverge. We take a Gaussian
defined by the mean value $\mu_1$ and variance $\sigma$, 
\begin{equation}
\mathcal{W}(T_c)=\frac{1}{\sqrt{2\pi}\sigma}{\mathrm e}^{-(T_c-\mu_1)^2/2\sigma^2}
\end{equation}
and
a Cauchy (i.e., Lorentzian) 
distribution defined by the mean value $\mu_2$ and the width $\gamma$, 
\begin{equation}
\mathcal{W}(T_c)=\frac{\gamma}{\pi[\gamma^2+(T_c-\mu_2) ^2]}
\end{equation}
Throughout the paper the same parameters $\mu_1=\mu_2=1$, $\sigma=0.1$ and 
$\gamma=0.08$ will be adopted. The parameters $\sigma$ and $\gamma$ are chosen so
that the resistivity curves obtained with the Gaussian and Cauchy distribution   
have the same slopes close to the transition when space correlations are absent (Fig. \ref{UniformComparison}).

As recalled in Sec. \ref{RRN}, one of the findings of Ref. [\onlinecite{CGBC}] was 
that {\it in the absence of space correlations} EMT performs remarkably well 
in reproducing the exact results of the RRN, irrespective of
the specific $T_c$ distribution. As an illustration, we report in Fig. \ref{UniformComparison} 
the resistivity curves of a uniform system obtained with the distributions here considered:
The Gaussian (blue curves) and the Cauchy distribution (green curves). One notices 
that the resistance curves obtained with EMT match very well the exact numerical 
solutions.
Thus, in the absence of correlations, EMT provides 
a useful tool to investigate the $\rho(T)$ dependence. 
In particular, the theory states that the resistivity of a RRN vanishes 
as soon as the weight of superconducting bonds $w_s$ exceeds $1/(1+\alpha)$, where 
$\alpha$ is the connectivity of the system. In the present case of a uniform two 
dimensional lattice, $\alpha=1$ and the percolation threshold at the transition yields $w_s=1/2$. 
Referring to the resistance curves in Fig. \ref{UniformComparison}, one 
notices that the critical temperature of the system is $T_{\alpha}=1$, coinciding 
with the mean value of the (symmetric) distributions where half of the bonds 
have become superconducting. 
Thus, we correctly reproduce the predicted value of $w_s=1/2$. At high temperature, 
the heavier tail of the Cauchy distribution results in a greater number of bonds 
with switched-off resistance.  The resistivity for the Cauchy distribution remains 
smaller until the temperature $T_{\alpha}=1$ is reached where 
$w^{\mathrm C}_s=w^{\mathrm G}_s=1/2$, the superscripts $C,G$ referring henceforth 
to the Cauchy and Gaussian distribution, respectively.

To give a quantitative picture of the resistivity close to $T_{\alpha}$ one can take 
the derivative of Eq. (\ref{solution}) defined in the framework of EMT:
\begin{equation}
\rho'_{em}(T)=(1+\alpha)\rho_0\mathcal{W}(T_c=T),
\end{equation}
for $T\geq T_{\alpha}$,
which shows that the slope of the resistivity is proportional to the value 
of the distribution at equal temperatures. Combining this equation with 
the previous result $w_{s}(T_{\alpha})=1/2$, where the distribution has a maximum,
one concludes that it is impossible to obtain ``tails'' in the present case.
In fact, due to the bell-shape of the distributions the slope of the $\rho(T)$ 
is maximal at the mean value and thus also at the critical temperature of the 
system. 
In the same line of reasoning, one realizes that 
a tailish behavior can be obtained even for a 
spatially uncorrelated distribution, in the very specific case of a symmetric 
bimodal distribution of $T_c$'s (see Ref. [\onlinecite{CGBC}]). Here, we disregard this possibility and focus 
instead on the issue of space correlations.

\subsection{Diffusion Limited Aggregation}
\label{DLA}
\begin{figure}[htbp]
\begin{center}
\includegraphics[scale=0.265]{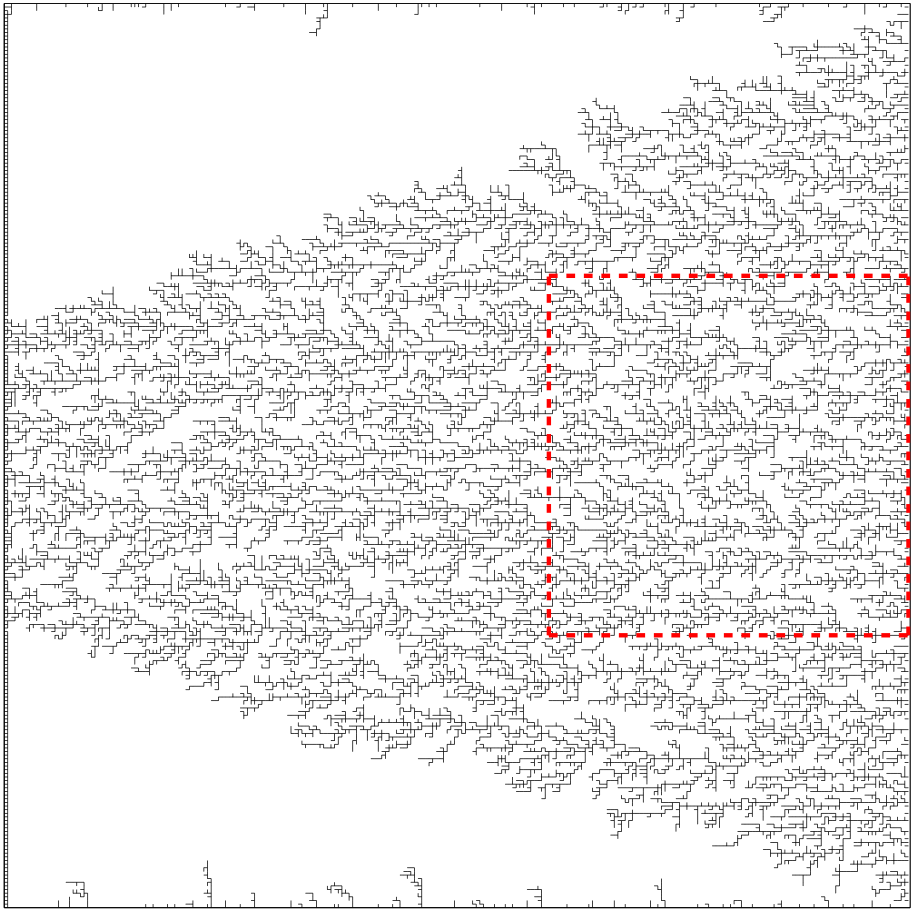}
\caption{Spatial distribution of superconducting bonds obtained through diffusion 
limited aggregation. The red square indicates the $100\times100$ lattice which is used for the calculations.}
\label{DLACluster}
\end{center}
\end{figure}

The first fractal-like cluster we implement is obtained through a simple 
growth process, generated by Brownian motion and known as 
\emph{diffusion limited aggregation} (DLA). Its construction is quite simple:
A particle is released at the left edge of a 2D lattice and let 
diffuse to the right. More precisely, the particle moves one bond to the right 
and then with equal probability one bond up or down. This procedure is iterated
until the particle stops, as soon as it reaches the top, bottom or right edge, where
it sticks. Then, other particles are launched in sequence and halted either 
when reaching one of the three edges or a bond already occupied by one of the 
previously diffused particles. The cluster obtained in a $250\times250$ square 
lattice after diffusing $50000$ particles is defined by the bonds where the particles 
sticked. Due to a saturation at the left edge, the total number of superconducting bonds 
only amounts to about $25000$. 
Once this large cluster is obtained, we select a $100\times100$ sub-lattice as shown in 
Fig. \ref{DLACluster} and perform the calculations for this smaller cluster. In this way, we 
try to model a more physical case where the fractal covers the whole sample.
So henceforth, we consider a $100\times100$ lattice, where
only bonds belonging
to the cluster are assigned a critical temperature $T_c$. 
The other bonds form a
resistive background and are assigned the resistivity $\rho_0$ at all temperatures.

\begin{figure}[htbp]
\begin{center}
\includegraphics[scale=0.21]{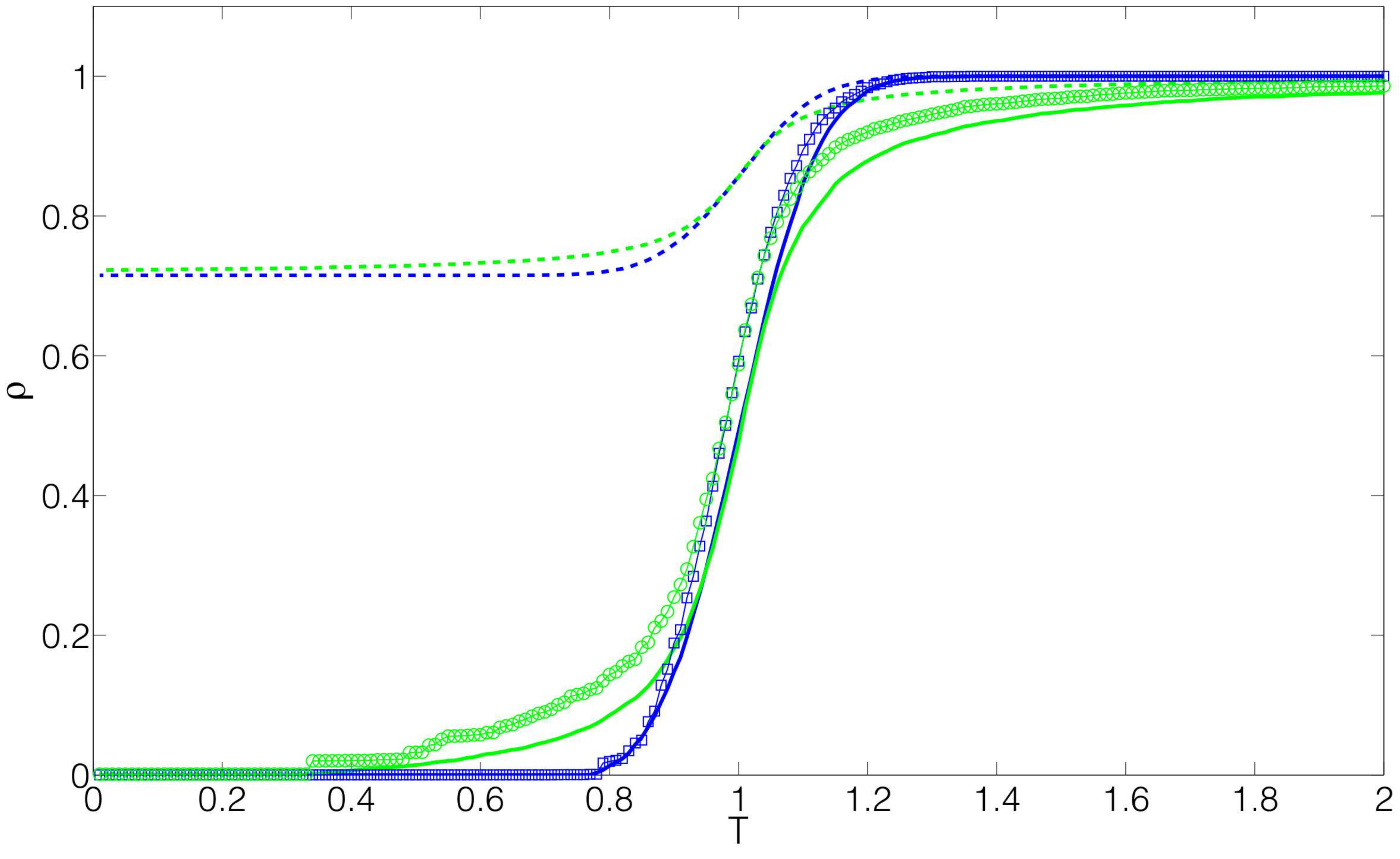}
\caption{(Color online)  Resistance curves of diffusion limited aggregation cluster. 
The blue curves correspond to a Gaussian distribution with $\mu=1,\sigma=0.1$
and the green to a Cauchy distribution with $\mu=1,\gamma=0.08$. 
The markers report the thermal evolution of the resistance obtained 
by exact numerical calculations, the dashed lines by EMT on the system and the full lines by EMT on the cluster.}
\label{DLAResistanceCurve}
\end{center}
\end{figure}

Fig. \ref{DLAResistanceCurve} reports the resistance curves calculated on the 
$100\times100$ DLA cluster. 
We remark that for $T\gtrsim1$ the curves for the Gaussian and Cauchy distribution are rather similar.
Lowering the 
temperature, the system starts to exhibit a stronger dependence on the specific 
statistics. In the Gaussian case, a percolating path is already formed 
around $T_{\alpha}=0.78$, whereas for the Cauchy distribution one has to go as low 
as $T_{\alpha}=0.33$ for the system to become fully superconducting.
Examining the formation of the superconducting cluster, one finds that 
at $T_{\alpha}$, the fraction of superconducting bonds are $w_s^{G}=0.99$ 
and $w_s^{C}=0.97$ respectively. So practically all the bonds have to be 
superconducting in order for the phase transition to occur.

The reason for this strong condition lies in the very small set of percolating 
paths, i.e., the \emph{effective} connectivity of the cluster is very close to zero. In other words, 
the cluster has a rather marked one-dimensional character and few missing 
superconducting bonds are enough to prevent the system from percolating.
To give a quantitative estimate of the effective  connectivity of the cluster 
we define the following 
measure (inspired by Eq. (\ref{w_s}) defined in the EMT framework),
\begin{equation}\label{alpha}
\alpha^{eff}=\frac{1}{w_s(T_{\alpha})}-1.
\end{equation}
For instance, 
comparing $w_s^{U}(T_{\alpha})=0.5$ and $w_s^{G, DLA}(T_{\alpha})=0.99$, where
the superscripts $U$ and $DLA$ refer to the case of a uniform and DLA 
superconducting cluster, respectively, one realizes 
that the statistical weight of the superconducting bonds clearly depends on the 
geometry of the cluster and, in particular close to the critical temperature $T_{\alpha}$, on its
effective connectivity. 
In addition, we shall show below that taking $\alpha^{eff}_{DLA}=1/0.99-1=0.01$ the EMT 
{\it restricted to the cluster} produces $\rho(T)$ curves close to the exact 
results.  
Therefore, even though space correlations do not enter explicitly, 
equation (\ref{alpha}) still provides a good estimate for an effective connectivity, once 
the relevant space correlations are accounted for by restricting EMT to the cluster.  
A second effect of the small set of percolation paths is a strong dependence on 
the specific realization of the distribution. Since the number of bonds in the 
percolating cluster is at most of order $N^{D}$ (where $D$ is the fractal dimension of the cluster),
two different realizations of the same 
distribution might be quite different.
This effect is particularly severe for the Cauchy distribution,
which has larger deviations from the mean. Since the Gaussian distribution, having
small deviations from the mean, produces a superconducting cluster with an effective connectivity 
$\alpha_{DLA}^{eff}=0.01$, we assume this value as representative of the appropriate connectivity of the 
underlying DLA cluster.
On the other hand, the realization of the Cauchy distribution of 
critical temperatures on the cluster in Fig. \ref{DLACluster} is such that a 
percolating path forms at finite temperature but, calculations of $\rho(T)$ 
for different realizations of $\mathcal W(T_c)$
show that on average the phase transition does not occur.  
This result can be understood by noticing that the Cauchy distribution has 
rather wide tails so that $w_s^{C}(T\leq0)\approx 0.025$. Thus the fraction of
bonds, which never become superconducting is substantial and, owing to the 
low connectivity of the DLA cluster (requiring $w^{DLA}_{s}=0.99$), it may well 
happen that non-superconducting 
bonds prevent full percolation to be established down to zero temperature.

This effect makes it quite apparent that on correlated clusters, the low values 
of connectivity (like $\alpha^{eff}_{DLA}=0.01$) renders the system very sensitive to 
the statistics of the critical temperatures.

Fig. \ref{DLAResistanceCurve} also shows that the EMT, when applied to the
whole RRN (which is formed both by the fractal-like cluster and its complementary 
background), clearly misses the correct temperature dependence of the 
resistivity. According to its construction, outlined in Sec. \ref{RRN}, this 
mean-field-like theory neglects all spatial correlations and this discrepancy 
is expected. In addition, we remark that the decrease of $\rho$ with $T$ is a finite-size 
effect: As soon as we introduce spatial correlations, the cluster dimension $D$ becomes smaller 
than $2$, and the statistical weight of superconducting bonds is smaller (or equal)
than $N^D/N^2$ which tends to zero for $N\rightarrow\infty$.

One can try to circumvent the shortcomings of EMT by evaluating it 
on the bonds belonging to the DLA cluster only. With this adjustment, 
the relevant space correlations are effectively incorporated and
EMT with the effective connectivity reproduces 
quite accurately the main features of the exact solution: The regular behavior 
for $T>\mu$ and the stronger dependence on the distribution at low temperatures, 
due to the cluster's low connectivity. 
However, there is an overall shift from the restricted EMT solution with respect to the exact calculation. 
At high temperatures, where only very few bonds have become superconducting, 
it is not favorable to force the current to flow through many resistive bonds simply to reach
a few superconducting bonds.
Therefore, the current still flows
through the system in parallel (only locally perturbed by the few superconducting bonds present).
By decreasing the temperature the superconducting cluster increases and consequently, 
also the amount of current going through it.
Close to the transition the shift then nearly vanishes because the resistive background
ceases to play an important role. More precisely, 
the current is essentially carried by one or a few large 
quasi-percolating paths, which only need few resistors to switch off in order to 
fully percolate. For these paths, it is immaterial whether they are embedded in a 
two-dimensional (markers in Fig. \ref{DLAResistanceCurve}) or simply a quasi 
one-dimensional system (full lines in Fig. \ref{DLAResistanceCurve}).


\subsection{Symmetrized Random Walk}\label{SRW}
The topology of the DLA cluster considered in the previous section is characterized by few
backbones (forming connected paths between the two vertical edges of the lattice) and many dangling branches.
Now, we explore a different situation in which the fractal-like geometry does not contain 
any dangling branches.

\begin{figure}[htbp]
\begin{center}
\includegraphics[scale=0.34]{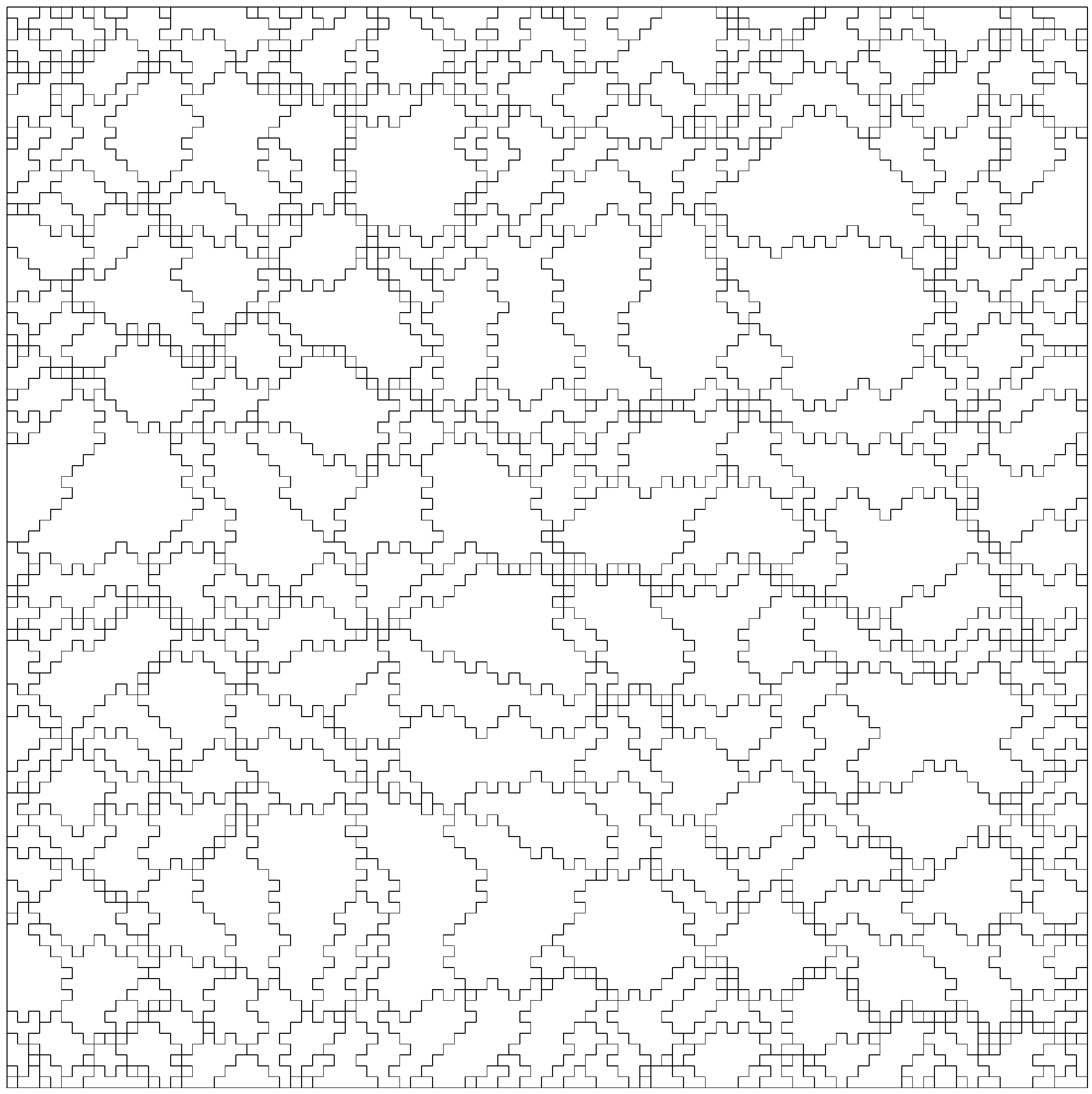}
\caption{Spatial distribution of superconducting bonds obtained through a 
symmetrized random walk on a $100\times100$ lattice.} 
\label{SRWCluster}
\end{center}
\end{figure}

We modify the construction scheme of the DLA in 
the following way: Instead of keeping only the final position of the diffused 
particles, their entire trajectory is incorporated into the cluster. In addition, 
the particles are free to cross the paths of previously diffused particles. 
Launching $50$ particles from each of the four edges and letting them diffuse 
perpendicular to the initial edge, a cluster of the form shown in 
Fig. \ref{SRWCluster} is obtained. 

\begin{figure}[htbp]
\begin{center}
\includegraphics[scale=0.21]{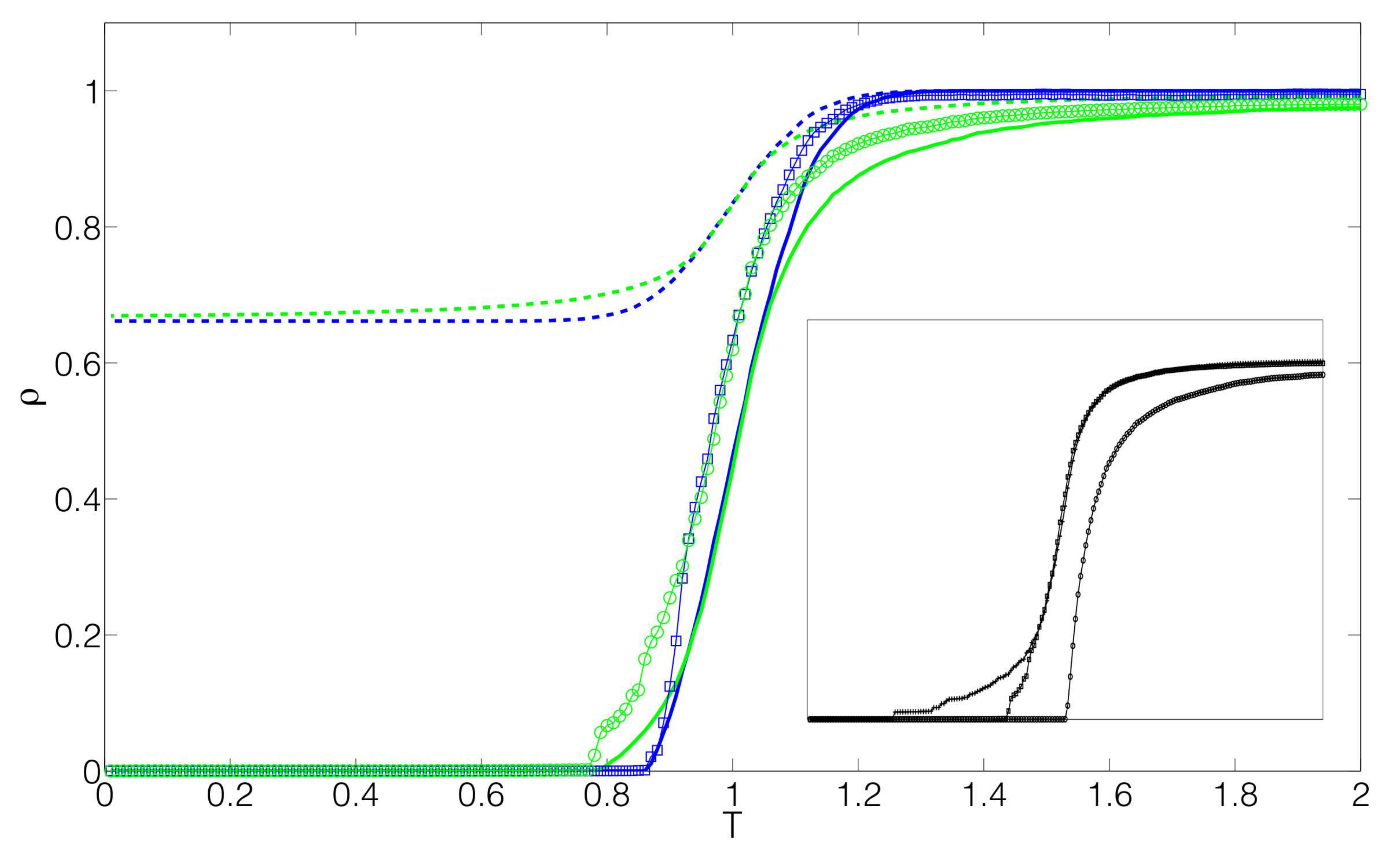}
\caption{(Color online)  Main panel: Resistance curves of symmetrized random walk cluster. 
The blue curves correspond to a Gaussian distribution with $\mu=1,\sigma=0.1$
and the green to a Cauchy distribution with $\mu=1,\gamma=0.08$. 
The markers report the thermal evolution of the resistance obtained 
by exact numerical calculations, the dashed lines by EMT on the system and the full lines by EMT on the cluster. The inset shows a comparison of the exact resistivity curves obtained with a Cauchy distribution ($\mu=1,\gamma=0.08$) for the uniform (circles), the DLA (pluses) and the SRW (squares) cluster, respectively. }
\label{SRWResistanceCurve}
\end{center}
\end{figure}

As it was the case for the DLA, the resistance curves of the \emph{symmetrized 
random walk} (SRW) shown in Fig. \ref{SRWResistanceCurve} are very similar 
for $T\gtrsim0.9$, the small difference originating simply from the difference in 
the distributions. Looking at the superconducting cluster formed at $T=0.9$, 
one finds that the weights of superconducting bonds are $w_s^{G}=0.84$ 
and $w_s^{C}=0.79$, respectively. Interestingly, this difference has no effect 
on the resistivity. In fact, at this temperature, the size of connected 
superconducting regions is of the order of $10$ bonds for both distributions. In 
this regime, having a few superconducting bonds more or less does not produce 
a noticeable difference in the resistivity. 

This behavior clearly changes as one approaches the transition temperature. 
For $T$ slightly larger than $T_{\alpha}$ the current is carried by long-range superconducting 
regions which lack only few bonds to fully percolate. Inspection of the superconducting
cluster growth reveals that the percolation threshold is reached as soon as 
$w_s^{G}(T_{\alpha}^G)=0.92$ and $w_s^{C}(T_{\alpha}^C)=0.90$. 
The closeness of these two values is again an indication that the threshold weight $w_s(T_{\alpha})$
only depends on the ``geometry" (i.e. its effective connectivity) of the underlying cluster and not
on the specific $T_c$ distribution. 
While the threshold 
is reached at $T_{\alpha}^G=0.86$ in the first case, the heavy tail of the Cauchy 
distribution obliges the system to go as low as $T_{\alpha}^C=0.76$ in the 
second case. 

\begin{figure}[htbp]
\begin{center}
\includegraphics[scale=0.16]{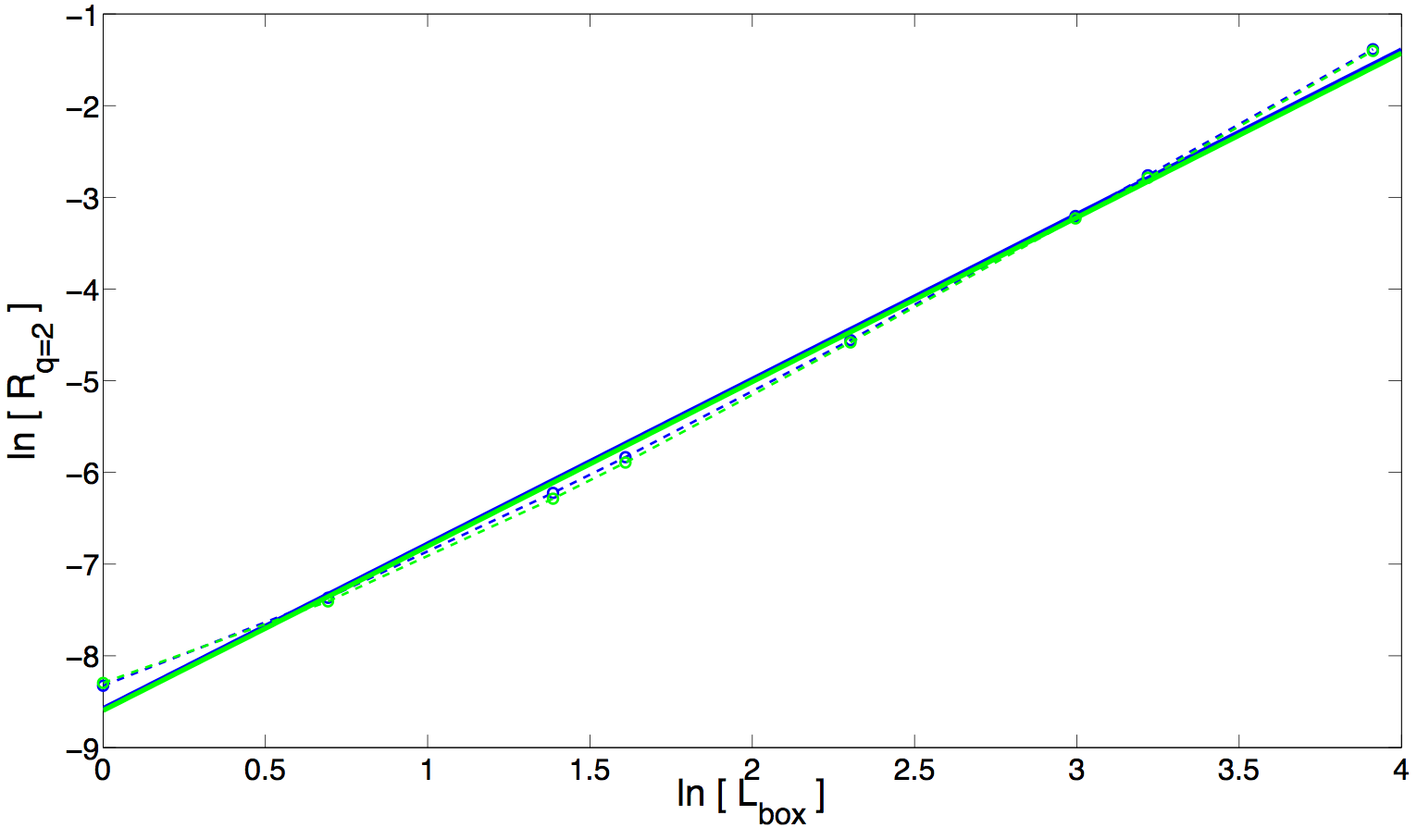}
\caption{(Color online)  Determination of fractal dimension using the box-counting method: The lattice is subdivided into $N_{box}=(N/L_{box})^2$ boxes of size $A_{box}=L_{box}\times L_{box}$. For each box $k$, we calculate the ratio $\mu_k=N^k_{bond}/N_{c}$, between the number of bonds belonging to the portion of the cluster inside this box $N^k_{bond}$ and the total number of bonds belonging to the cluster $N_{c}$. The quantity $R_q=\sum_{k=1}^{N_{box}}\mu_k^q$ is then related to the box size according to $R_q=L_{box}^{D(q-1)}$, where $D$ is the fractal dimension. The figure shows a log-log plot with $L_{box}$ varying from $1$ to $N/2$ (circles). The full lines are linear regressions of $\text{ln}\,R_{q=2}\propto D\,\text{ln}\,L_{box}$; the DLA cluster (green curve) and the SRW cluster (blue curve) have both slope $D=1.8$.}
\label{DLA_SRW_q2}
\end{center}
\end{figure}

These considerations on $w_s^{U}$, $w_s^{DLA}$ and $w_s^{SRW}$ suggest that the high-temperature behavior is 
mainly determined by the dimensionality of the cluster. In fact, 
coarse-graining analysis using the box-counting method\cite{Halsey} shown in Fig. \ref{DLA_SRW_q2} reveals that  
the dimension of the clusters is $D^{DLA}\approx D^{SRW}\approx 1.8$. 
This is reflected in the congruence of $\rho^{DLA}(T)$ and $\rho^{SRW}(T)$ for $T>0.9$,
as well as in their leftward shift with respect to $\rho^{U}(T)$ (with $D^U=2$), as is shown in 
the inset of Fig. \ref{SRWResistanceCurve}.


In contrast, the calculation of the connectivity at the transition point reveals 
that $\alpha^{eff}_{SRW}\approx0.1$, which corresponds to a nearly one-dimensional cluster. 
So, despite its 2D appearance, the low-temperature region, being 
more sensitive to the effective connectivity than to the dimensionality, behaves as
quasi-one-dimensional.
As far as the comparison between the EMT and the numerical exact results is 
concerned, the conclusions are analogous to the discussion given for the DLA 
cluster.

To develop a quantitative understanding of $\rho(T)$ close to the transition, we 
turn to an analytic approach based on a coarse-grained picture of the cluster. 
We consider the random  walk cluster as a uniform system made up by large 
conducting segments containing $M\gg 1$ bonds. In the framework of EMT the 
percolation threshold is reached when $1/2$ of the segments are superconducting. 
One can reformulate the condition in terms of the single bonds; 
in order for a large segment to become superconducting, all its single bonds have to be superconducting, i.e.,
\begin{equation}
p^M=1/2
\end{equation}
where $p$ is the probability for a single bond to be superconducting.
Close to the global critical point, the resistivity can be approximated by
\begin{equation}
\rho(T\gtrsim T_c)=C\rho_0\left(\frac{1}{2}-p^M\right),
\end{equation}
with $C>0$. The purpose of the model is to analyze the possible occurrence of a 
tail in the resistivity curve, i.e., the behavior of the slope of $\rho(T)$
at $T\approx T_c$:
\begin{eqnarray}\label{rhoprime}
\frac{d\rho}{dT}\Big|_{T\approx T_c}&=&-Mp^{M-1}p'
\end{eqnarray}  
where $C\rho_0$ was set to $1$ for simplicity and $p'=dp/dT$.
Since $T\approx T_c$ and $M\gg1$  the following approximations can be made,
\begin{eqnarray}
p^M&\approx &\frac{1}{2}\\
p&\approx&\exp\left[-\frac{1}{M}\ln(2)\right]\approx 1-\frac{1}{M}\ln(2)\\
&\Rightarrow& M=\frac{\ln(2)}{1-p}
\end{eqnarray}  
Inserting the two expressions into Eq. \ref{rhoprime}, we get an expression for 
the derivative of the resistivity as a function of the single bond probability close 
to the critical point,
\begin{eqnarray}
\frac{d\rho}{dT}\Big|_{T\approx T_c}&=&-\frac{1}{2}\ln(2)\frac{p'}{p(1-p)}\label{Condition}\\
&\approx&-\frac{1}{2}\ln(2)\frac{p'}{1-p}\\
&=&\frac{1}{2}\ln(2)\frac{\mathcal{W}(T_c)}{1-\int_{T_c}^{\infty}
\mathcal{W}(\tilde{T}_c)d\tilde{T}_c}.
\end{eqnarray}   
In terms of the distribution the condition $p\approx1$ (see Eq. (\ref{Condition})) means that the 
leading behavior is obtained taking the limit 
$\mathcal{W}(T_c\rightarrow-\infty)$ (or, more precisely, 
$\mu-T_c\gg\sigma,\gamma$; we point out that the critical temperature stays
finite). This implies that the slope 
of the resistivity is proportional to the ratio of two small numbers. For the 
Gaussian distribution, the ratio is
\begin{eqnarray}\label{WGauss}
&&\frac{\mathcal{W}(T_c)}{1-\int_{T_c}^{\infty}\mathcal{W}(\tilde{T}_c)d\tilde{T}_c}
\approx\frac{\frac{1}{\sqrt{2\pi}\sigma}{\mathrm e}^{-x^2}}{1-\left(1-
\frac{{\mathrm e}^{-x^2}}{\sqrt{\pi}|x|}\right)}
\nonumber\\ 
&&=\frac{|x|}{\sqrt{2}\sigma}\approx\frac{|T_c-\mu_1|}{2\sigma^2},
\end{eqnarray}   
where the short-hand notation $x=(T_c-\mu_1)/\sqrt{2}\sigma$ was used.
For the Cauchy distribution, the ratio is
\begin{eqnarray}\label{WCauchy}
&&\frac{\mathcal{W}(T_c)}{1-\int_{T_c}^{\infty}\mathcal{W}(\tilde{T}_c)d\tilde{T}_c}
\approx\frac{\frac{\gamma}{\pi[\gamma^2+(T_c-\mu_2) ^2]}}{1-
\left(1-\frac{\gamma}{\pi T_c}\right)}
\nonumber\\
&&\approx\frac{1}{|T_c-\mu_2|}.
\end{eqnarray} 
One observes that the slope for the Gaussian distribution depends linearly on the deviation from the 
critical temperature, while in the case of the Cauchy distribution the dependence 
is inversely proportional. The evaluation of the expressions (\ref{WGauss}) and 
(\ref{WCauchy}) with the results of the SRW yields 
$\rho'(T_c^{G})\approx\rho'(T_c^{C})\approx0.15$, which is smaller than the 
exact results given in Fig. \ref{SRWResistanceCurve}. The reason for this 
mismatch lies in the assumption $M\gg 1$, while in the present case $M\approx 5$
. Nevertheless, the calculations highlight the strong influence of 
the distribution for low-dimensional structures
with $M\gg1$.  


To investigate the SRW and illustrate the analytic approach 
presented above, we briefly consider a cluster obtained through coarse graining. 
The cluster contains every fifth column and fifth row of the original network. In 
this way, one obtains a uniform grid equivalent to the uniform case of a 
$N/5\times N/5$ lattice where each bond is made up by $5$ bonds of the original 
network.  

\begin{figure}[htbp]
\begin{center}
\includegraphics[scale=0.21]{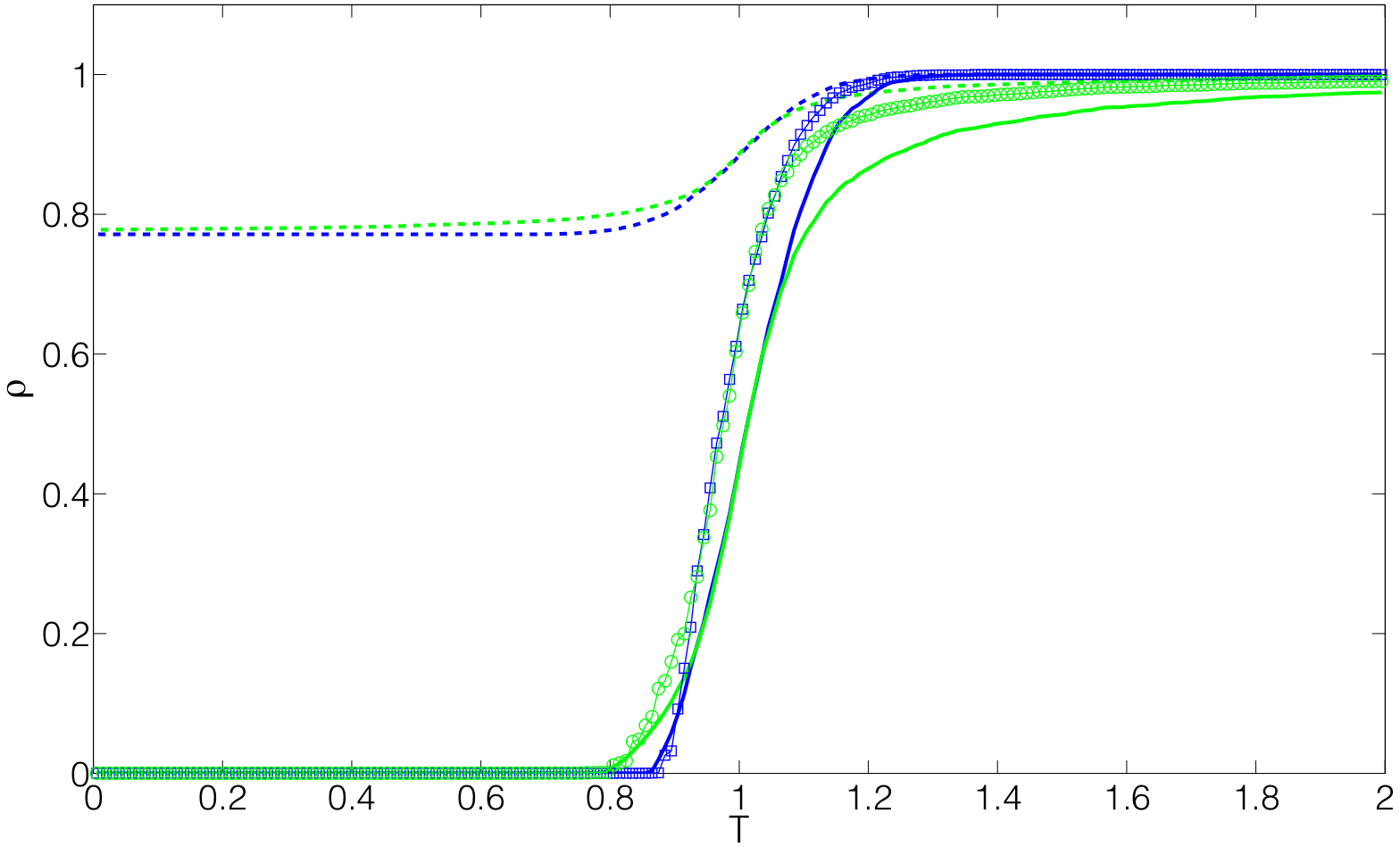}
\caption{(Color online)  Resistance curves of coarse grained cluster.
The blue curves correspond to a Gaussian distribution with $\mu=1,\sigma=0.1$
and the green to a Cauchy distribution with $\mu=1,\gamma=0.08$. 
The markers report the thermal evolution of the resistance obtained 
by exact numerical calculations, the dashed lines by EMT on the system and the full lines by EMT restricted to the cluster.}
\label{CoarseGrained}
\end{center}
\end{figure}

Comparison of Fig. \ref{CoarseGrained} and Fig. \ref{SRWResistanceCurve} reveals 
a marked resemblance between the $\rho(T)$ curves for the random walk and the 
coarse grained cluster. This shows that the regular coarse-grained cluster indeed captures
the main features of the strongly disordered SRW cluster.

\section{Conclusions}   
Our analysis is purely phenomenological in character: We embed a given
fractal-like structure in a purely metallic environment, we impose a given $T_c$ distribution on it,  
and then we calculate the resulting resistivity curve. While, as we already mentioned, this approach
lacks any (possible, but not mandatory) microscopic connection between the geometrical and the $T_c$
distribution, we separately access the distinct effects of these two ingredients.

As a first result,
we would like to point out that the geometrically
correlated character of the superconducting clusters considered here greatly degrades
the performance of the EMT mean-field-like approach.
According to our experience in spatially uncorrelated disorder\cite{CGBC}, this failure 
has nothing to do with the strength of disorder or the relative density of the superconducting
and the non-superconducting bonds, but is a mere result of space correlations. 
On the other hand, we showed that it is possible to find a good qualitative agreement by restricting EMT to the cluster. 
In this case, one needs to introduce an effective connectivity (see Eq. (\ref{alpha})).

Calculating the standard (local) connectivity
given by $\alpha=\langle z \rangle/2-1$, where $\langle z \rangle$ is the average number of nearest neighbors,
we obtain $\alpha_{DLA}=0.38$ and $\alpha_{SRW}=0.15$.
The comparison with $\alpha^{eff}_{DLA}=0.01$ and $\alpha^{eff}_{SRW}=0.1$ reveals that 
the effective connectivity is very different 
from the locally defined connectivity.
This is quite natural since we are considering transport phenomena, which are ruled by paths 
connecting distant regions of the system. In other words,
the presence of regions with large internal connectivity, is rather immaterial for transport if this 
regions are weakly connected to one another. 

We also find that the effective connectivity is an intrinsic property of the cluster and is independent
of the $T_c$ distribution.
This result is quite natural because upon reducing $T$, the 
$T_c$ distribution only determines how ``fast" the resistances are switched off. However, this is 
irrelevant as far as the amount of superconducting bonds needed to percolate is concerned.

The main outcome of our analysis is that the effects of space correlations render
the superconducting transition strongly dependent on the disorder distribution, and in
particular on its low-temperature asymptotics. This is not surprising: As soon as 
percolation happens through a quasi one-dimensional path, the system needs to explore
(almost) the entire $T_c$ distribution, and thus becomes very sensitive to its low-temperature
part.
This 
effect can be made particularly clear within the analytic approach of Sec. \ref{SRW}, 
where the slope of the resistivity at $T_c$ was shown to depend in inverse ways 
on the distance of $T_c$ from the mean of the Cauchy or Gaussian distribution.

After the detailed analysis of the various effects of space correlations and $T_c$ 
statistics we come back to the initial question of whether and which of these 
aspects is relevant in explaining the tailish behavior of $\rho(T)$ in oxide interfaces. 
Based on the above results a possible ingredient is indeed a fractal-like structure of superconducting 
bonds. Due to its high dimensionality (i.e. larger than a purely one-dimensional cluster), the 
cluster consists of enough superconducting bonds to cause a linear decrease of the resistance with 
a relatively high slope. Close to the percolation threshold, however, the crucial property determining 
the shape of the transition is not the dimensionality but the effective connectivity. There, the 
quasi-one-dimensional behavior of the DLA and SRW clusters results in a more or less pronounced tail of 
$\rho(T\gtrsim T_{\alpha})$. The shape of the tail is then controlled (to some extent) by the low-temperature asymptotics
of the distribution.

Our analysis and approach are therefore a useful tool to explore the physical mechanisms at work in 
the superconducting oxide interfaces, where a precise fitting of the resistance curves provides 
indications on both the intra-grain mechanisms leading to a local $T_c$ and to the inter-grain 
connection ruling the establishing of the global superconducting state via the percolative 
``chaining'' of superconducting regions {\it on the fractal support} (see Sec. \ref{framework} for 
the specific meaning of the word ``percolative" in the present model). 
The intragrain superconductivity
manifests itself in the high-temperature region of the resistivity, where the restivity starts 
to bend down when {\it isolated} superconducting regions are created. In this case the $T_c$ distribution 
and the relation between its width and the high temperature resistivity might provide informations on 
the intra-grain pair formation (Cooper pairs in the presence of quenched impurities,\cite{finkelstein}
disordered bosonic preformed pairs,\cite{fisher} glassy superconducting 
transition.\cite{ioffe,feigelman10}) On the other hand, near the global $T_c$, the second  
percolative phenomenon is ruled by the geometric character (in particular by the effective connectivity) of the
superconducting cluster, and might be informative about the origin of the
inhomogenous metal-superconductor structure with coexisting metallic and superconducting islands.


\vskip 0.5truecm 
\par\noindent 
{\bf Acknowledgments.} 
We are indebted with N. Bergeal, J. Biscaras, C. di Castro, B. Leridon, J. Lesueur, and J. Lorenzana,   
for interesting discussions and useful comments.  S.C., C. C., and M.G. acknowledge 
financial support from ``University Research Project'' of the ``Sapienza''
University n. C26A115HTN.

\end{document}